\documentclass[prb,twocolumn,floatfix,showpacs]{revtex4}
\usepackage{graphicx,epsfig}
\usepackage{amsmath,amssymb,bm}
\usepackage{mathrsfs}

\begin{document}

\title{Indirect forces between impurities in one-dimensional quantum liquids}
\author{P.~W\"achter}
\affiliation{Institut f\"ur Theoretische Physik, Universit\"at G\"ottingen, 
Friedrich-Hund-Platz 1, D-37077 G\"ottingen, Germany}
\author{V.~Meden}
\affiliation{Institut f\"ur Theoretische Physik, Universit\"at G\"ottingen, 
Friedrich-Hund-Platz 1, D-37077 G\"ottingen, Germany}
\author{K.~Sch\"onhammer}
\affiliation{Institut f\"ur Theoretische Physik, Universit\"at G\"ottingen, 
Friedrich-Hund-Platz 1, D-37077 G\"ottingen, Germany}

\begin{abstract}

We investigate the indirect interaction between two isolated impurities 
in a Luttinger liquid described by a microscopic lattice model.
To treat the electron-electron interaction $U$ the functional 
renormalization group method is used. For comparison we also 
study the $U=0$ case. We find that for a wide range of impurity
parameters the impurity  interaction $V_{12}$ as a function of 
their separation $r$ oscillates with decaying amplitude between 
being attractive and repulsive. For half-filling of the band and in a
crossover regime between weak and strong impurities the interaction
becomes purely attractive. For $U=0$ and independent of the impurity
strength the amplitude of the interaction energy falls off as $1/r$. 
For $U>0$ the decay for small separations and weak to intermediate 
impurities is governed by a $U$ dependent exponent larger than $-1$, 
which crosses over to $-1$ for large $r$. The crossover scale depends 
on the impurity strength and $U$. We present simple pictures which 
explain our results in the limits of weak and strong impurities. We 
finally also consider attractive interactions $U<0$. 

\end{abstract}
\maketitle     

\section{Introduction}
\label{intro}

The physical properties of interacting one-dimensional (1D) electron 
systems differ substantially from the generic Fermi liquid behavior of 
higher-dimensional systems, offering a whole variety of interesting 
new effects.\cite{Giamarchi04} Replacing the Fermi liquid concept 
of quasi-particles, the Luttinger liquid (LL) phenomenology has proved 
to capture the low-energy physics of a large class of 
models.\cite{Schoenhammer05} The low lying excitations of the system 
are no longer described by quasi-particles, but rather by collective 
density excitations. Accordingly, even the presence of a single 
isolated impurity can have strong effects on the properties of the 
system.\cite{LutherPeschel,Mattis,ApelRice,KaneFisher,Furusaki0}

For the case of two isolated impurities placed in a LL 
the linear conductance has been investigated 
in detail.\cite{KaneFisher1,Furusaki2,Nazarov,Polyakov,VM6,Enss05} 
Other aspects have not yet been discussed to this extent. 
Here we study the indirect interaction between two impurities 
mediated by the electrons. 
A particular promising candidate for measuring such forces are 
impurities in fermionic ``atomic quantum wires'' realized with 
ultracold gases.\cite{Moritz} 
For a continuum model of noninteracting electrons with delta
impurities it was shown that the impurity
interaction $V_{12}$ as a function of the impurity separation $r$ oscillates 
between being attractive and repulsive. 
Its magnitude decays as $1/r$. 
The period of the oscillation is $\pi/k_F$, with $k_F$ being
the Fermi momentum.\cite{Recati05}
The effect of a repulsive 
interaction was then taken into account using a field theoretical 
effective low-energy model and considering the two limits of 
weak and strong bare impurities.\cite{Recati05,Fuchs06} 
The weak impurity case was analyzed using linear response theory while for
strong impurities methods developed in the
context of quantum Brownian motion were applied.\cite{Willi} 
In addition, for strong repulsive 
electron-electron interactions boundary conformal field theory was 
used to study the impurity interaction.\cite{Caux03} 

Here we 
supplement the earlier studies.
We consider a microscopic lattice model, focus on small to intermediate 
electron-electron interaction and apply a method which is 
nonperturbative in the strength of the impurities. This allows us to
study the crossover from the weak to the strong impurity behavior.
For the two limits simple pictures of the observed physics are presented. 
Some emphasis is put on the question whether the impurity interaction as a
function of $r$ continues to oscillate also in the presence of
electron-electron interaction.\cite{Caux03,Recati05,Fuchs06} We 
briefly discuss the case of attractive interactions.

The method of choice for our calculations is the recently 
developed functional renormalization group (fRG), which has turned 
out to be a powerful tool in the study of low-dimensional 
interacting electron systems.\cite{fRGgeneral,lecturenotes} 
We use an approximate 
truncation scheme that by comparison to
exact and numerical results was shown to provide a good approximation 
for small to intermediate electron-electron interaction. It was 
used for interacting systems of up to $10^7$ lattice
sites.\cite{Andergassen04,Enss05}  
 
The paper is organized as follows. In Sec.\ \ref{model} we  
introduce our model and in Sec.\ \ref{frg} give a brief outline of the 
fRG procedure used. In Secs.\ \ref{nonintcase} and \ref{intcase} 
we present our results for the $r$ dependence of the 
impurity interaction considering 
noninteracting and interacting electrons, respectively. We establish 
contact to the former calculations. Finally, in Sec.\ \ref{summary} 
our findings are
summarized.

\section{The model}
\label{model}

As our microscopic model we use the lattice model of spinless fermions
with nearest-neighbor hopping and 
nearest-neighbor interaction on a large but finite number $N$ of 
lattice sites. To suppress 
the effect of the boundaries the interacting chain is coupled to
noninteracting semi-infinite leads via adiabatic contacts. They 
are realized by varying the interaction smoothly across the 
two contacts from zero to the bulk value $U$.\cite{Enss05} We are
mainly interested in the half filled band case.  

The kinetic part of the Hamiltonian reads
\begin{equation}
\label{hamkin}
H_{\textnormal{kin}}=- t \sum_{j=-\infty}^{\infty}\left(c^\dagger_{j+1}c_j+
\textnormal{H.c.}\right)\quad\textnormal{,}
\end{equation}
where $c_j^\dagger$ and $c_j$ denote the fermionic
creation and annihilation operators on lattice site $j$. 
We set $t=1$ and use
the hopping as the unit of energy. In addition, the lattice spacing is
set to one.
The electrons are assumed to interact on the bonds between the 
sites $[1,N]$ via a spatially dependent nearest-neighbor 
interaction
\begin{equation}
\label{hamint}
H_{\textnormal{int}}=\sum_{j=1}^{N-1}U_{j,j+1}\left(n_j-\frac{1}{2}\right)
\left(n_{j+1}-\frac{1}{2}\right)\quad,
\end{equation}
defining the interacting wire of interest. 
The operators on the right-hand side of Eq.\ (\ref{hamint}) 
are the local density operators $n_j=c_j^\dagger c_j$  
shifted by $-1/2$ to assure that the interacting part of the system is
half filled. To turn on the interaction 
smoothly we use the function
\begin{equation}
U_{j,j+1}=U\frac{\arctan[s(j-j_s)]-\arctan[s(1-j_s)]}
{\arctan[s(\frac{N}{2}-j_s)]-\arctan[s(1-j_s)]}
\end{equation}
with $s=\frac{1}{4}$ and $j_s=56$ for the left part 
$[1,\frac{N}{2}]$ of the interacting wire and a similar function 
for the right part. In combination with the semi-infinite 
leads this provides us with a wide region in the center of the interacting 
wire, which has a constant interaction strength $U$.  
Hardly any effects from the contacts can be 
detected.\cite{Andergassen04,Enss05}

\begin{figure}[tb]
\begin{center}
\includegraphics[width=0.45\textwidth,clip]{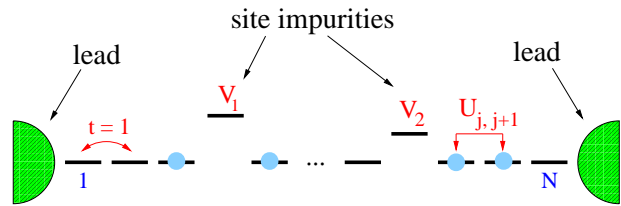}
\end{center}
\vspace{-0.6cm}
\caption[]{(Color online) Schematic plot of the lattice model 
with site impurities; in the case of hopping impurities $t\neq 1$ 
for two bonds\label{fig1}}
\end{figure}

In the region of constant $U$ we place the two impurities 
whose interaction we want 
to study. We mainly consider site impurities, modeled by 
the Hamiltonian ($V_\alpha > 0$)
\begin{equation}
\label{hamimp}
H_{\textnormal{imp}}=\sum_{\alpha=1}^{2}V_\alpha n_{j_\alpha}\quad,
\end{equation}
and sketched in Fig.\ \ref{fig1},
but also use hopping impurities described by ($0 < t_\alpha <1$)
\begin{equation}
\label{hamimpp}
H_{\textnormal{imp}}'=-\sum_{\alpha=1}^{2}(t_\alpha-1)
\left[c^\dagger_{j_{\alpha}+1}c_{j_\alpha}+\textnormal{H.c.}\right]\quad.
\end{equation}
For both types of impurities the impurity separation $r$ is defined as
$r= j_2-j_1$.

The homogeneous model $H=H_{\textnormal{kin}}+H_{\textnormal{int}}$ 
with constant interaction between all bonds (that is also in the
semi-infinite leads) can be solved exactly 
via a Bethe ansatz\cite{YangYang} and exhibits LL behavior for all 
fillings and all $U$ except for half-filling with 
$|U|\geq 2$.\cite{Haldane80} In the half filled case, the relation 
between the LL parameter $K$, which later on will become important, 
and the interaction 
strength $U$ can be given in a closed form\cite{YangYang,Haldane80} 
\begin{equation}
\label{exK}
K^{-1}=\frac{2}{\pi}\arccos\left(-\frac{U}{2}\right)\quad.
\end{equation}

\section{Functional renormalization group}
\label{frg}

\subsection{General setup}

The general formulation of the fRG for the generating functional 
$\Gamma$ of the one-particle irreducible $n$-particle vertices 
starts by introducing an infrared cutoff $\Lambda$  in the 
free propagator leading to a $\Lambda$ dependent $\Gamma^\Lambda$.
We here use a sharp energy cutoff.\cite{Andergassen04}
By differentiating with respect to $\Lambda$ one can derive 
an exact infinite hierarchy of coupled differential flow equations for 
the vertex functions.\cite{fRGgeneral,lecturenotes}

The adaption of this general scheme to our present problem, namely 
to a spinless inhomogeneous LL at temperature $T=0$, and the 
approximations involved in deriving a closed set of equations   
for the self-energy
are described in great detail in Ref.\ \onlinecite{Andergassen04}. 
The self-energy is approximated as frequency independent
$\Sigma^\Lambda(i \omega) \to \Sigma^\Lambda$. Defining 
$\tilde{G}^\Lambda(i\omega) = [G_0(i\omega) - \Sigma^\Lambda]^{-1}$,
with the Green function $G_0$ obtained from the kinetic 
part Eq.\ (\ref{hamkin}) of the Hamiltonian,
the flow equations for the spatial (Wannier basis) matrix elements read
\begin{align}
\label{flowsys1}
\partial_\Lambda\Sigma_{j,j}^\Lambda&=-\frac{1}{\pi}\sum_{r=\pm 1}
U^\Lambda_{j,j+r} \, \mbox{Re} \, \tilde{G}_{j,j+r}^\Lambda(i \Lambda)
\; ,
\\\label{flowsys2}
\partial_\Lambda\Sigma_{j,j\pm1}^\Lambda&=
\frac{1}{\pi}U_{j,j+1}^\Lambda  \, \mbox{Re} \, 
\tilde{G}_{j,j\pm1}^\Lambda(i \Lambda) \; ,
\end{align}
with 
\begin{align}
\label{flowsys3}
U^\Lambda_{j,j+1}&=\frac{U_{j,j+1}}{1+\left(\Lambda-\frac{2+\Lambda^2}
{\sqrt{4+\Lambda^2}}\right)\frac{U_{j,j+1}}{2\pi}}\quad .
\end{align}
The 2-particle vertex was parametrized by a static
nearest-neighbor 
interaction of strength $U^\Lambda_{j,j+1}$ which implies that the
self-energy is a tridiagonal matrix. 
The flow equation (\ref{flowsys3})  of the 2-particle vertex 
is the especially simple form for half-filling, derivable from the general 
formula at arbitrary filling.\cite{Andergassen04}
The fRG flow leads 
from $\Lambda=\infty$ down to $\Lambda=0$, where the original 
system is recovered. 
At the end of the flow  
$\Sigma^{\Lambda=0}_{j,j'}$ present the  
frequency independent approximation for the self-energy. 
This approximation scheme to the full hierarchy of flow equations 
was successfully used to study various aspects 
of inhomogeneous LLs.\cite{Andergassen04,VM6,Enss05}

For our lattice model the matrix elements of the 
inverse of the full Green function at scale $\Lambda$, 
$[\tilde{G}^\Lambda]^{-1}(i \Lambda)$ 
for $j,j' \in [1,N]$ are given by
\begin{align}\label{greensfunc}\nonumber
& [\tilde{G}^\Lambda]_{j,j'}^{-1}(i \Lambda) = 
i \Lambda\delta_{j,j'}+(\delta_{j,j'+1}+\delta_{j,j'-1})\\
&-\frac{1}{2}\left(i \Lambda-\sqrt{(i \Lambda)^2-4}
\right) \left(\delta_{j,j'}\delta_{j,1}+\delta_{j,j'}
\delta_{j,N}\right) - \Sigma_{j,j'}^\Lambda \,.
\end{align}
The second to last term effectively accounts for the 
semi-infinite leads as an additional contribution to the 
self-energy.\cite{Enss05}

The initial values at cutoff 
$\Lambda_0 \gg 1$ 
of the fRG flow equations (\ref{flowsys1}) and (\ref{flowsys2}) 
in the case of site impurities are given by\cite{Andergassen04}
\begin{align}\label{initialvalues1}
\Sigma_{j,j}^{\Lambda_0}&=\sum_{\alpha=1}^2V_\alpha\delta_{j,j_\alpha}
\; ,\\\label{initialvalues2}
\Sigma_{j,j\pm1}^{\Lambda_0}&= 0 \; ,
\end{align}
whereas for hopping impurities one finds
\begin{align}
\label{initialvalueshop1}
\Sigma^{\Lambda_0}_{j,j}&=0 \; ,\\
\label{initialvalueshop3}
\Sigma^{\Lambda_0}_{j,j\pm 1}&=
-\sum_{\alpha=1}^2(t_\alpha-1)\delta_{j,j_\alpha}
\; .
\end{align}
 
\subsection{0-particle vertex and grand canonical potential}
\label{0PVandGCP}

Earlier papers using the fRG for 1D systems mainly focused on the 
calculation of the self-energy and observables which can directly be
computed from it. In the present paper we are interested in the 
grand canonical potential (GCP) from which the energy of the
indirect impurity interaction can be obtained.  
The connection between the GCP and the 0-particle vertex  
(0PV) is established by replacing 
the normalizing grand canonical partition function $\mathscr{Z}$ 
in the thermodynamical average by the grand canonical 
partition function $\mathscr{Z}_0$ of the free system, that is the 
system without interaction and impurities. This keeps the vertex 
functions of order greater or equal one unaltered and merely changes 
the physical meaning of the 0PV as can be seen by looking at its 
definition in terms of the generating functional $\Gamma$:\cite{Negele}
\begin{equation}
\Omega=\Gamma(\{\phi\},\{\overline{\phi}\})\vert_{\phi=\overline{\phi}=0}=-\ln
\left(\mathscr{Z}\right)+\ln\left(\mathscr{Z}_0\right)\quad,
\end{equation}
with the external source fields $\phi$ and $\overline{\phi}$.
Thus, the 0PV describes the difference between the GCP of the full 
system and the one of the free system.

For our sharp energy cutoff the flow equation of the 
0PV reads\cite{fRGgeneral,lecturenotes}
\begin{equation}
\label{omstart}
\partial_\Lambda\Omega^\Lambda=\frac{1}{2\pi}\sum_{\omega=\pm\Lambda}
\textnormal{Tr}\ln\left[\mathbf{1}-\Sigma^\Lambda(i \omega)
  G_0(i \omega)\right] \, e^{i \omega \eta}
\quad,
\end{equation}
were the trace and the matrix structure refer 
to the quantum numbers and we explicitly introduced the 
convergence factor 
$e^{i \omega\eta}$, which is usually suppressed.
The limit $\eta \searrow 0$ has to be taken at the end 
of the calculations. 
The validity of Eq.\ (\ref{omstart}) is not restricted to the 
approximation scheme used here, which leads to a frequency
independent self-energy.

Since in a numerical solution the flow necessarily starts from a 
large but finite cutoff $\Lambda_0$, we have to take into account 
the flow from $\Lambda=\infty$ to $\Lambda_0$ in the initial 
values. For the self-energy and the 2-particle vertex this 
procedure is described in Ref.\ \onlinecite{Andergassen04}.
In this large $\omega$ limit on can replace $\Sigma^\Lambda(i \omega)$
by its frequency independent part $\Sigma^\Lambda$ [even if a more
sophisticated truncation than the one leading to Eqs.\
(\ref{flowsys1})-(\ref{flowsys3}) is used].  
Approximating $G_0$ by its high frequency behavior 
$G_{0;j,j'}(i \omega)\approx \delta_{j,j'}/
(i \omega)$ and expanding the logarithm we obtain with
$\Omega^{\Lambda=\infty}=0$
\begin{equation}
\Omega^{\Lambda_0}\approx\frac{1}{\pi}
\int_{\Lambda_0}^\infty \textnormal{d} \Lambda
\frac{\sin{(\eta \Lambda)}}{\Lambda} \; \textnormal{Tr} \,
\Sigma^\Lambda \; .
\end{equation}
With the large $\Lambda$ behavior of
$\Sigma^\Lambda$ (see Ref.\ \onlinecite{Andergassen04}) 
\begin{align}
\Sigma_{j,j'}^\Lambda \approx \Sigma_{j,j'}^{\Lambda=\infty} +
\frac{1}{\pi} \sum_{l} I_{j,l;j',l} \int_{\Lambda}^\infty
\textnormal{d} \Lambda'  \frac{\sin{(\eta \Lambda')}}{\Lambda'} \; ,
\end{align}
where $I_{i,j;k,l}$ is the bare antisymmetrized interaction, 
one obtains performing the integrals over $\Lambda$ and $\Lambda'$ in
the limit $\eta \searrow 0$
\begin{equation}
\Omega^{\Lambda_0}\approx\frac{1}{2}  \textnormal{Tr} \, W+\frac{1}{8}
·\sum_{j,j'}  I_{j,j';j,j'}
\; .
\end{equation}
Here $W$ denotes all single-particle terms of the Hamiltonian not
taken into account in the free propagator $G_0$.\cite{foot1} 
For our problem this yields the initial values
\begin{equation}
\Omega^{\Lambda_0}=\frac{V_1+V_2}{2}-\frac{1}{4}\sum_{j=1}^{N-1}U_{j,j+1}
\; ,
\end{equation}
for site impurities and
\begin{equation}
\Omega^{\Lambda_0}=-\frac{1}{4}\sum_{j=1}^{N-1}U_{j,j+1} \; ,
\end{equation}
for hopping impurities.
After this step $\eta$ can be set to zero in Eq.\ (\ref{omstart}). 
Using the matrix identity 
$\textnormal{det}\,e^A=e^{\textnormal{Tr}A}$ Eq.\ (\ref{omstart}) 
simplifies to  
\begin{equation}
\label{flowsys0}
\partial_\Lambda\Omega^\Lambda =\frac{1}{\pi}\ln\left\vert
\frac{\det[\tilde{G}^\Lambda]^{-1}(i \Lambda)}
{\det G_0^{-1}(i \Lambda)}\right\vert \; .
\end{equation}

\section{Noninteracting case}\label{nonintcase}

\subsection{Analytical results}

In the noninteracting case, we compute the energy of the 
indirect interaction of two site impurities by evaluating the
partition function. For finite temperatures $T$ this gives
\begin{align}\nonumber
\frac{\mathscr{Z}}{\mathscr{Z}_{0}}&=\prod_{\omega_n}
\Bigl\{1-G_{0;j,j}(i \omega_n)\left[V_1+V_2\right]\\
&\phantom{=\prod_{\omega_n}\Bigl\{}+V_1V_2
\left[G_{0;j,j}^2(i \omega_n)-G_{0;j,j+r}^2(i \omega_n)\right]
\Bigr\}\quad,
\end{align}
where the product involves all Matsubara frequencies $\omega_n$ and  
$r \in {\mathbb N}$ denotes the distance between the two impurities. 
At half-filling (with chemical potential $\mu=0$) 
the Green function of the noninteracting and impurity free system is
given by 
\begin{equation}\label{greendisk}
G_{0;j,j+r}(z)=\frac{1}{\sqrt{z^2-4}}\left(-\frac{z}{2}+\frac{1}{2}
\sqrt{z^2-4}\right)^r \; .
\end{equation}

The interaction energy between the impurities $V_{12}(r)$ as function 
of their separation $r$ can be obtained as the difference of the 
GCP for infinitely separated impurities and the GCP for finite  
$r$.\cite{Recati05} This leads to
\begin{widetext}
\begin{equation}\label{WWergoWW}
V_{12}=-\frac{1}{\pi}\int_0^\infty \textnormal{d}\omega\ln
\left\vert1-\frac{V_1V_2G_{0;j,j+r}^2(i \omega)}{1-G_{0;j,j}(i \omega)
\left[V_1+V_2\right]+V_1V_2G_{0;j,j}^2(i \omega)}\right\vert\quad,
\end{equation}
\end{widetext}
where we took the limit $T \to 0$ and replaced the sum over 
Matsubara frequencies by an integral. 

In the limits of weak $V_1,V_2 \ll 1$ and strong $V_1 , V_2 \gg
1$ impurities and for sufficiently large $r$ 
(that is for $r \gg 1/k_F$; throughout  this work we will not be
interested in the behavior of $V_{12}$ at very small separations $r
\lesssim 1/k_F$)  
Eq.\ (\ref{WWergoWW}) with the Green function Eq.\ (\ref{greendisk})
can further be evaluated. For small $V_1,V_2 \ll 1$  we obtain
\begin{eqnarray}
\label{weakimplimit}
V_{12} = (-1)^{r+1} \frac{V_1 V_2}{4 \pi r} 
\end{eqnarray}
and for large $V_1,V_2 \gg 1$
\begin{eqnarray}
\label{strongimplimit}
V_{12} = \left\{ \begin{array}{cc}
\frac{\pi}{6 r} & \mbox{for $r$ even} \\
- \frac{\pi}{12 r} & \mbox{for $r$ odd.} 
\end{array} \right.
\end{eqnarray} 
In both limits the impurity interaction oscillates around zero 
with a periodicity of two lattice sites, which corresponds to
$\pi/k_F$ with the Fermi momentum $k_F=\pi/2$ at half filling, and decays
as $1/r$ with increasing impurity separation. For strong impurities 
the oscillation around zero is asymmetric, while it is symmetric 
for weak impurities. Furthermore, increasing the impurity strength 
from weak to strong  for fixed $r$ the impurity interaction 
changes from being attractive to being repulsive (even separations) 
and vice versa (odd separations).

\subsection{Numerical results}

\begin{figure}[tb]
\begin{center}
\includegraphics[width=0.4\textwidth,clip]{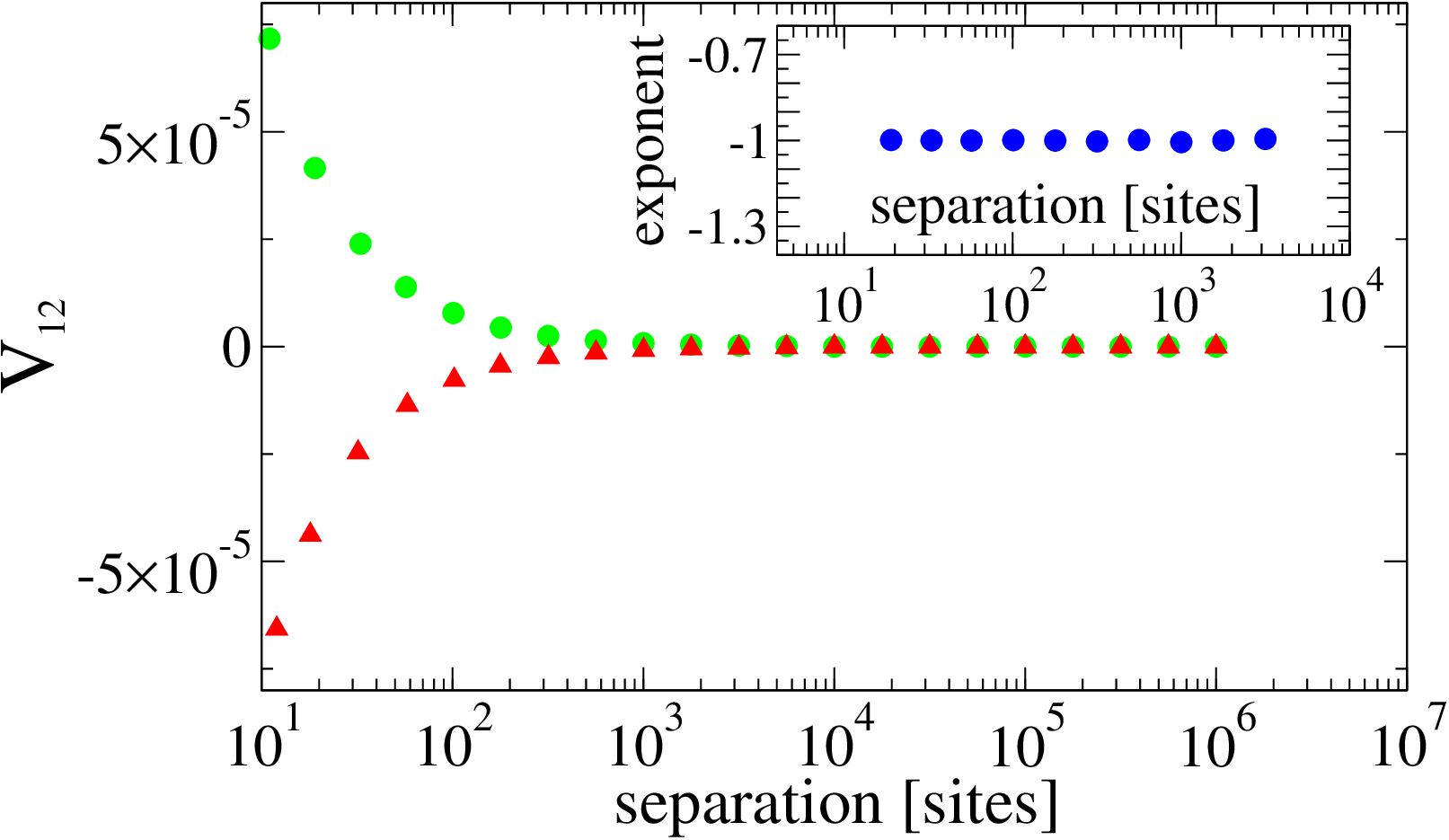}
\end{center}
\vspace{-0.6cm}
\caption[]{(Color online) Oscillations of the impurity interaction as 
function of the separation for $V_1=V_2=0.1$ in the noninteracting 
case. Triangles: even separations; circles: odd separations.
Inset: exponent of the decay.\label{oWWVp1}}
\begin{center}
\includegraphics[width=0.4\textwidth,clip]{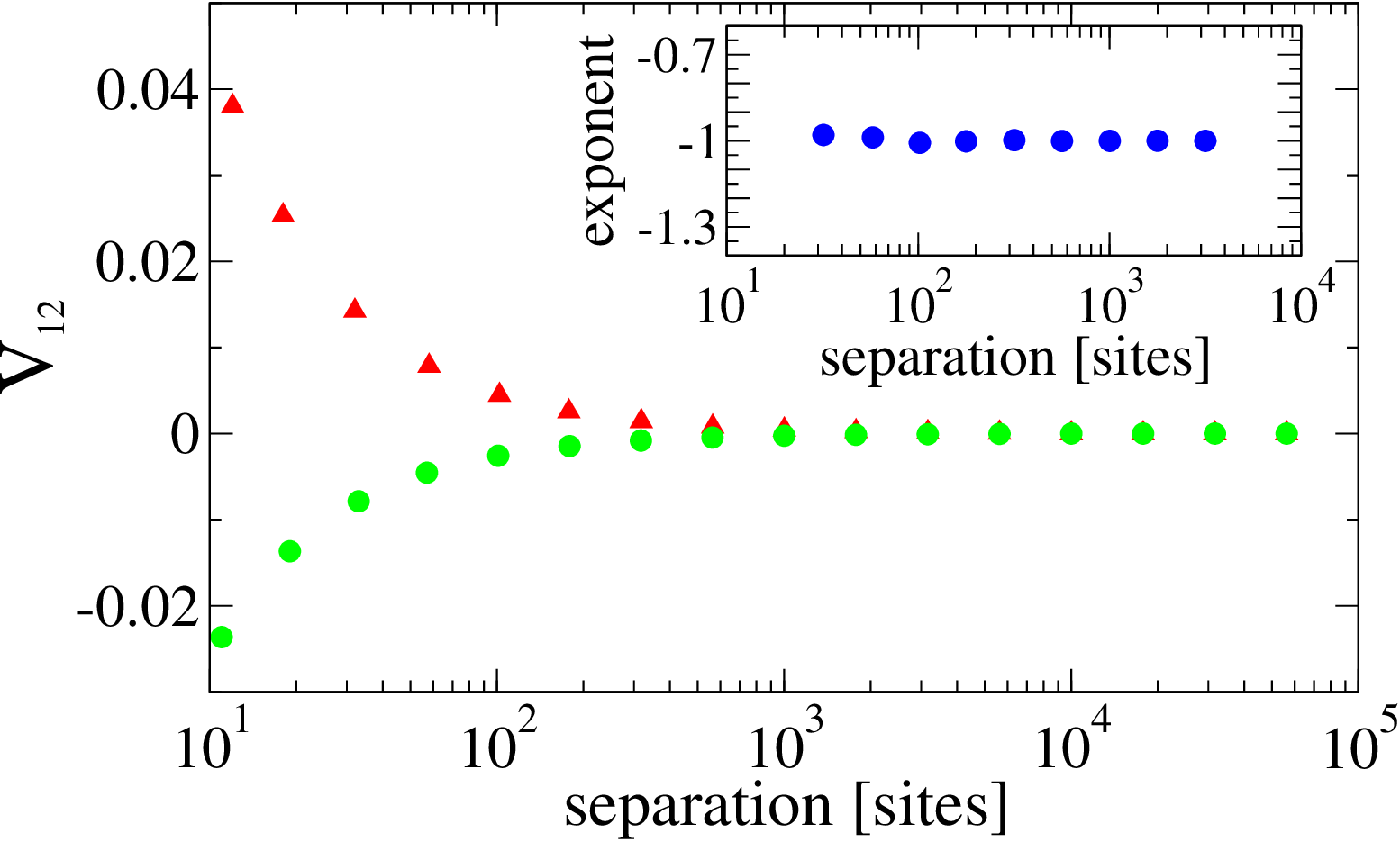}
\end{center}
\vspace{-0.6cm}
\caption[]{(Color online) The same as in Fig.\ \ref{oWWVp1} but 
for $V_1=V_2=30$.\label{oWWV30}}
\begin{center}
\includegraphics[width=0.4\textwidth,clip]{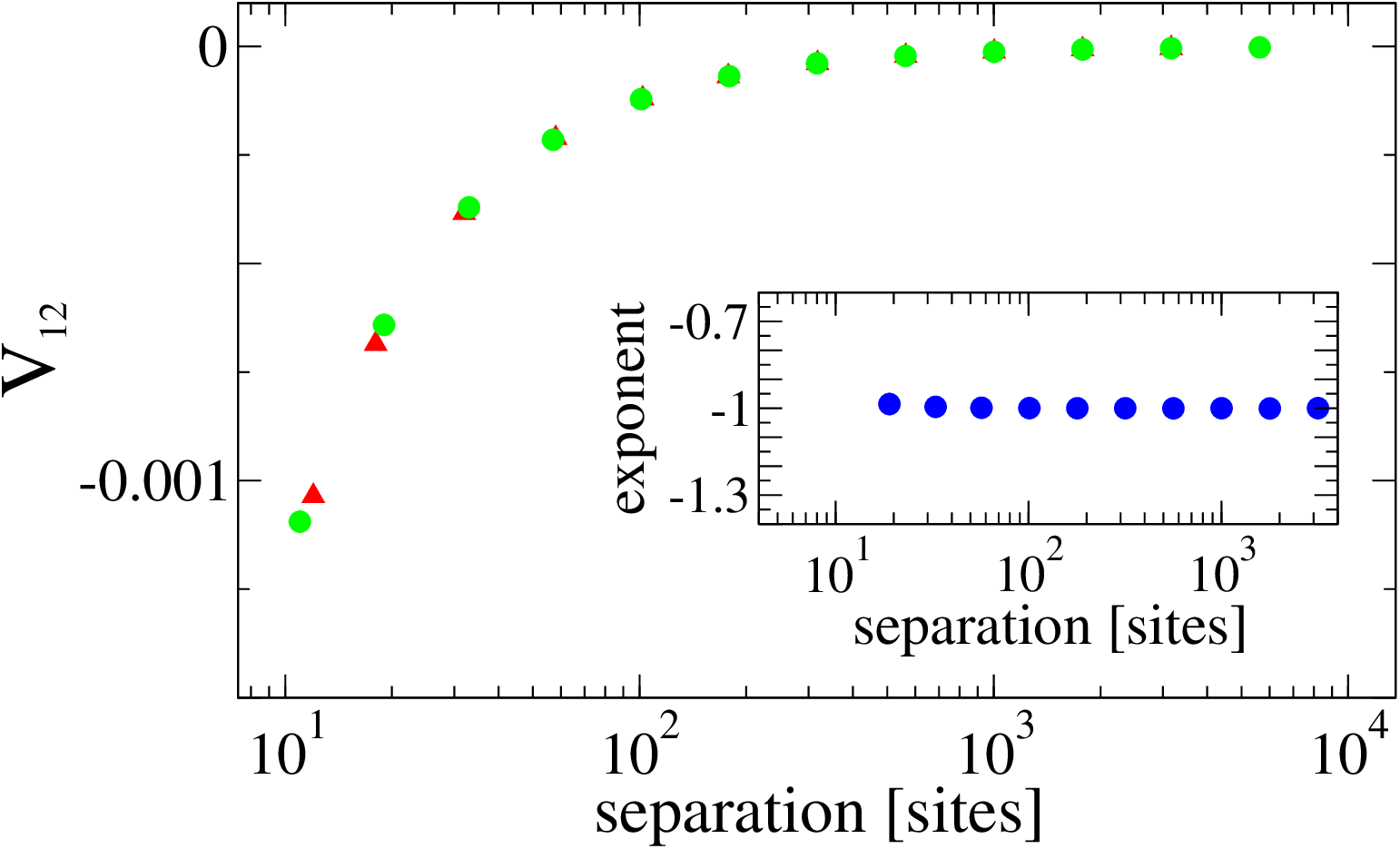}
\end{center}
\vspace{-0.6cm}
\caption[]{(Color online) Disappearance of oscillations for $V_1=1$ 
and $V_2=4$ in the case of noninteracting electrons. 
Triangles: even separations; circles: odd separations.
Inset: exponent of the decay.\label{oWWnooszi}}
\end{figure}

For weak and strong impurities the $r$-dependence of 
$V_{12}$ is shown in Figs.\ \ref{oWWVp1} and \ref{oWWV30},
respectively. To obtain these results  we inserted the Green 
function  Eq.\ (\ref{greendisk}) into Eq.\ (\ref{WWergoWW}) 
and performed the integral numerically (not relying on any
approximations). For simplicity we chose 
$V_1=V_2$. To cover a wide range of separations the data are 
shown on a logarithmic scale.
We always plot two subsequent separations. In the figures the 
triangles indicate a separation of an even number of lattice sites, 
while the circles symbolize odd separations. 

We can define an effective exponent of the decay of $V_{12}$ 
as the logarithmic derivative of the difference of the amplitudes 
with respect to the separation. This exponent is shown in the insets 
of Figs.\ \ref{oWWVp1} and \ref{oWWV30}.
Consistent with the asymptotic analytical results 
Eqs.\ (\ref{weakimplimit}) and (\ref{strongimplimit})   
the interaction decays with the inverse of the separation.
A $1/r$ decay is also found for impurities of intermediate strength as
exemplified in the inset of Fig.\ \ref{oWWnooszi} obtained for $V_1=1$ and 
$V_2=4$ (for a discussion of the main part of Fig.\ \ref{oWWnooszi}, 
see below).  

\subsection{Interpretation}

For weak impurities the $1/r$ decay and the symmetric
oscillations with period $\pi/k_F$
can be understood from the spatial dependence 
of the Friedel oscillations of the electron density induced by a 
single impurity.\cite{textbook} 
For distances sufficiently larger then $\pi/k_F$ 
the change of the density oscillates symmetrically
around zero with period $\pi/k_F$ and in 1D dies off as the inverse 
distance from the impurity. In lowest order perturbation theory in the 
impurity strength this leads to a potential of similar shape at the 
position of the second impurity and thus to the observed behavior 
of $V_{12}$. This argument can also be applied to hopping
impurities and we again [as in Eq.\ (\ref{weakimplimit})] 
obtain a symmetrically oscillating (attractive 
for even $r$, repulsive for odd $r$) $V_{12}$ which decays as 
$1/r$.  

Also for strong impurities a simple picture of the  
$r$-dependence of $V_{12}$ can be given. To this end we consider a
simplified model of a half filled tight-binding chain with 
  {\it open boundary conditions} which follows  in the limit of 
  a single infinitely strong impurity (site or hopping). A second 
  infinitely strong 
   impurity is placed at distance $r$ from the first one. The 
  ground state energy of this setup can easily 
  be computed analytically leading to Eq.~(\ref{strongimplimit}) 
  if the infinitely strong impurities are modeled as site impurities, 
  while it gives 
\begin{eqnarray}
\label{strongimplimithop}
V_{12} = \left\{ \begin{array}{cc}
-\frac{\pi}{12 r} & \mbox{for $r$ even} \\
 \frac{\pi}{6 r} & \mbox{for $r$ odd.} 
\end{array} \right.
\end{eqnarray} 
for hopping impurities. We thus find that for this type of impurities
the interaction $V_{12}$ is attractive (repulsive) for even (odd) 
separations in the limit of weak as well as of {\it strong}  
impurities.\cite{fuchsreasoning} Simple numerics shows that the sign
also does not change for hopping impurities of intermediate strength. 

This analysis reveals that the $1/r$ decay as well as the prefactors 
$\pi/6$ and $-\pi/12$ follow from an even simpler consideration. 
$V_{12}$ can directly be extracted from the $1/N$ correction $\Delta E$
of the groundstate energy of a tight-binding chain with open 
boundary conditions. A simple analytic calculation shows that 
$\Delta E$ is given by $\pi/(6 N)$ 
for odd $N$ and $-\pi/(12 N)$ for even $N$. For {\it even} 
$N$ the chemical potential $\mu=0$ lies between the last 
occupied and the first unoccupied state. In this case the results 
from (boundary) conformal field  theory can be used. Within this 
approach one obtains for a model with conformal charge $c=1$
\begin{eqnarray}
\label{bCFT}
\Delta E = -\pi \frac{v_c}{24 N} \; , 
\end{eqnarray}
where $v_c$ denotes the charge velocity.\cite{conformal} This result 
also holds for $U \neq 0$.   
For the tight-binding model at $U=0$ one finds $v_c=2$ 
and Eq.\ (\ref{bCFT}) leads to $\Delta E= -\pi/(12 N)$ obtained 
above by direct calculation. 
To deduce the impurity interaction from these considerations
for {\it hopping impurities} 
the chain length $N$ must  be substituted by $r$. 
For {\it site impurities} a length $N$ corresponds to an 
impurity separation $r=N+1$. This explains the 
interchange of ``even''
and ``odd'' in  Eqs.\ (\ref{strongimplimit}) and (\ref{strongimplimithop}).
The reason for the $1/r$ decay of $V_{12}$ for strong impurities is 
thus the appearance of $1/N$ corrections to the groundstate energy of
a finite system with open boundary conditions. 

Our results for  $V_{12}$ in the lattice
model are equivalent to the ones found in the noninteracting continuum 
model with $\delta$ impurities investigated in Refs.\
\onlinecite{Recati05} and \onlinecite{Fuchs06}. There a different
interpretation is given in the limit of strong impurities.

\subsection{Other fillings and fine tuned parameters}

We  here assumed the system to be half filled. Relaxing this 
does not affect the general behavior of the impurity interaction, 
namely the oscillation of the interaction energy decaying with the 
inverse separation. Merely the periodicity is altered. The behavior in
the two limits of weak and strong impurities can again be understood
within the two simple pictures presented above. 

We next describe a behavior which we exclusively observed for
site impurities and at half-filling. In a narrow crossover regime 
between the limits of strong and weak impurities the interaction 
becomes entirely attractive, but still oscillates around a decaying 
average value with amplitudes which scale as $1/r$ (not shown here). 
A peculiar behavior is found under the constraint $V_2 = 4/V_1$. 
For such fine tuned impurity parameters the oscillations  
completely disappear. The latter is depicted in the main part of 
Fig.\ \ref{oWWnooszi}. Note that the amplitude still decays as $1/r$ 
as shown in the inset.

\section{Interacting case}\label{intcase}

To compute the interaction energy $V_{12}$ between the impurities 
for $U \neq 0$, we numerically solve 
the flow equations (\ref{flowsys1})-(\ref{flowsys3}) and 
(\ref{flowsys0}) with 
the appropriate initial values. 
To isolate the energy of the impurity interaction, we evaluate 
the GCP for different configurations, namely for $V_1=V_2=0$ yielding 
the extensive part of the electronic interaction, 
$V_1=0$ and $V_2 \neq 0$ at site of interest and vice versa, finally 
the GCP of the full system with $V_1 \neq 0$ and $V_2 \neq 0$. 
The interaction energy $V_{12}$ is then given 
by 
\begin{align}\nonumber
V_{12}&=(\Omega-\Omega_{V_1=V_2=0})-(\Omega_{V_1=0\neq V_2}-\Omega_{V_1=V_2=0})\\
&\phantom{=}-(\Omega_{V_2=0\neq V_1}-\Omega_{V_1=V_2=0})\quad.
\end{align}
For the case of hopping impurities the same equation holds, if 
one replaces $V_\alpha$ by $t_\alpha$ and keeps in mind that the 
impurity free case corresponds to $t_1=t_2=1$.

For $U \neq 0$ we examine systems of $2.5 \times 10^5$ sites, 
placing the impurities in the 
interval $[22500,200000]$ well apart from the sections of the lattice 
in which the interaction is turned on and off. This enables us to 
calculate the interaction energy for impurity separations of up to 
$10^5$ lattice sites. 
We restrict ourselves to 
half-filling and first consider bulk interactions $U\in(0,2]$, 
which correspond 
to LL parameters $K\in[0.5,1)$ [see Eq.\ (\ref{exK})]. 

\subsection{Repulsive electron-electron interaction}

We focus on equal $V_{1}$ and $V_2$ but again verified that deviating
from this restriction does not qualitatively change our main 
conclusions. Further down we briefly comment on the behavior for fine
tuned impurity parameters as discussed at the end of the last section
for $U=0$. Then it will also become important to consider $V_1 \neq
V_2$. 

\begin{figure}[tb]
\begin{center}
\includegraphics[width=0.4\textwidth,clip]{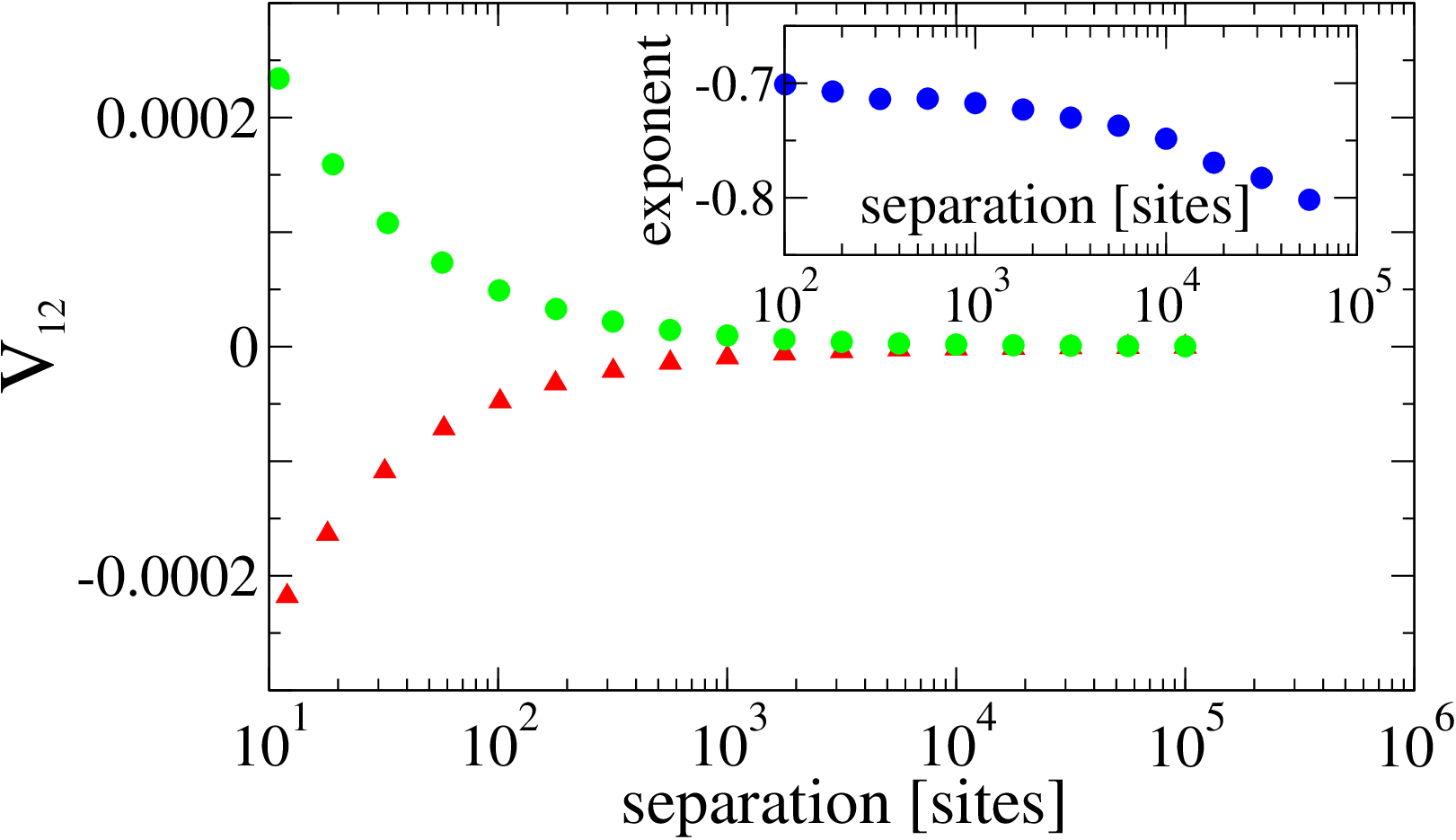}
\end{center}
\vspace{-0.6cm}
\caption[]{(Color online) Oscillations of the interaction energy as 
function of the separation for $V_1=V_2=0.1$ and interaction $U=0.5$. 
Triangles: even separations; circles: odd separations.
Inset: exponent of the decay.\label{WW1Vp1}}
\begin{center}
\includegraphics[width=0.4\textwidth,clip]{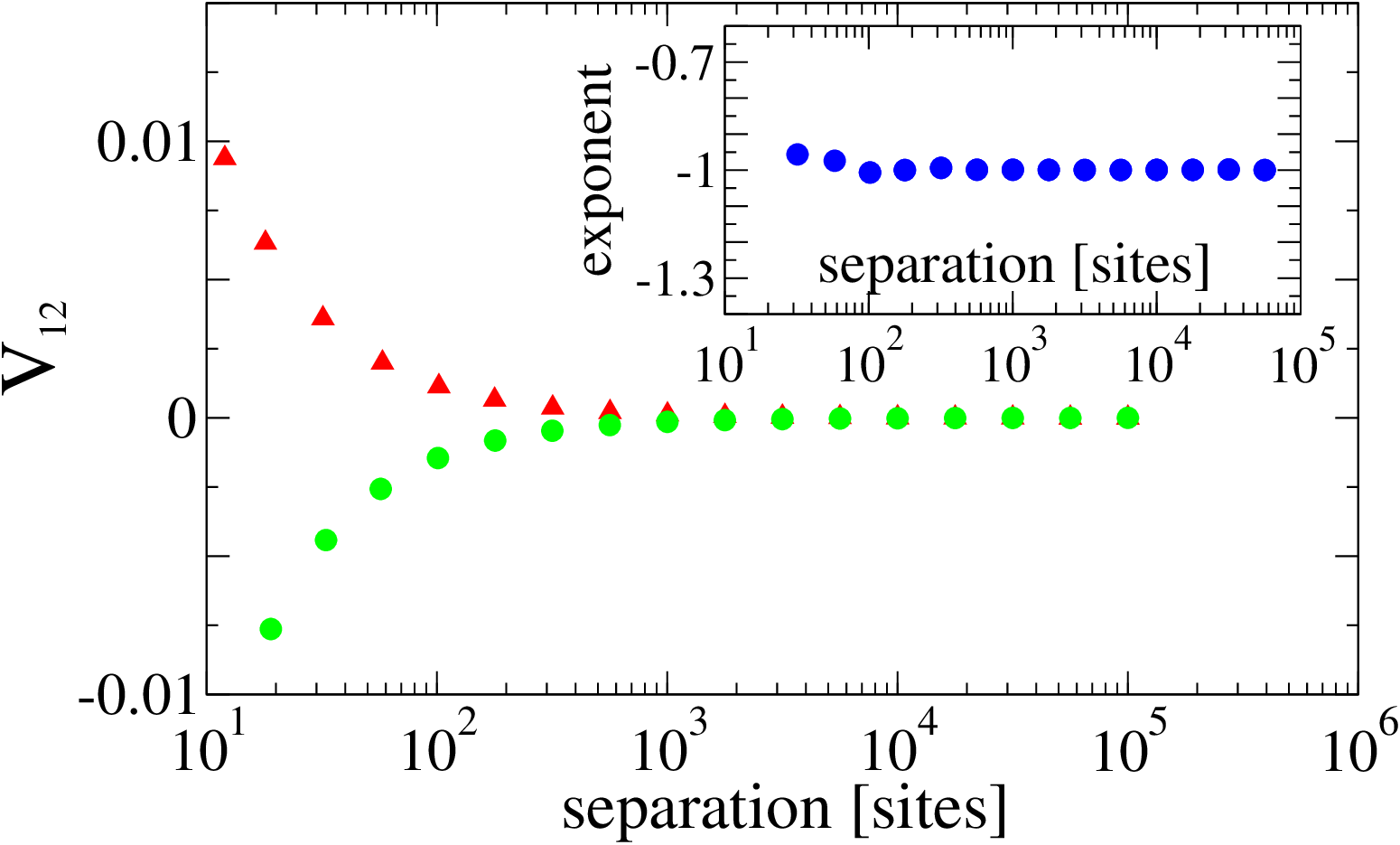}
\end{center}
\vspace{-0.6cm}
\caption[]{(Color online) The same as in Fig.\ \ref{WW1Vp1}, but for 
$V_1=V_2=10$. \label{WW1V10}}
\end{figure}

Figures \ref{WW1Vp1} and \ref{WW1V10} show the $r$ dependence 
of the interaction energy for weak and strong impurities. 
We find that the oscillation of the 
interaction with a periodicity of two lattice sites 
(corresponding to half-filling) is {\it robust} even in the presence of
the electron-electron interaction.\cite{Caux03,Recati05,Fuchs06}  
As for $U=0$ the impurity interaction for weak site impurities is 
repulsive for odd $r$ while it is attractive for even $r$. For strong
impurities the opposite holds.  Apart from the oscillation the 
interaction energy decreases as a function of $r$ and we extract the
``effective'' $r$-dependent exponent as in the last subsection. 
It is depicted in the insets of Figs.\ \ref{WW1Vp1} and \ref{WW1V10}. 
For strong impurities a clear $1/r$ decay is found. 
For weak to intermediate impurities as in Fig.\ \ref{WW1Vp1} the 
exponent is different from $-1$. It starts well above $-1$ for
  small $r$ and tends towards it with increasing separation. 

\begin{figure}[tb]
\begin{center}
\includegraphics[width=0.4\textwidth,clip]{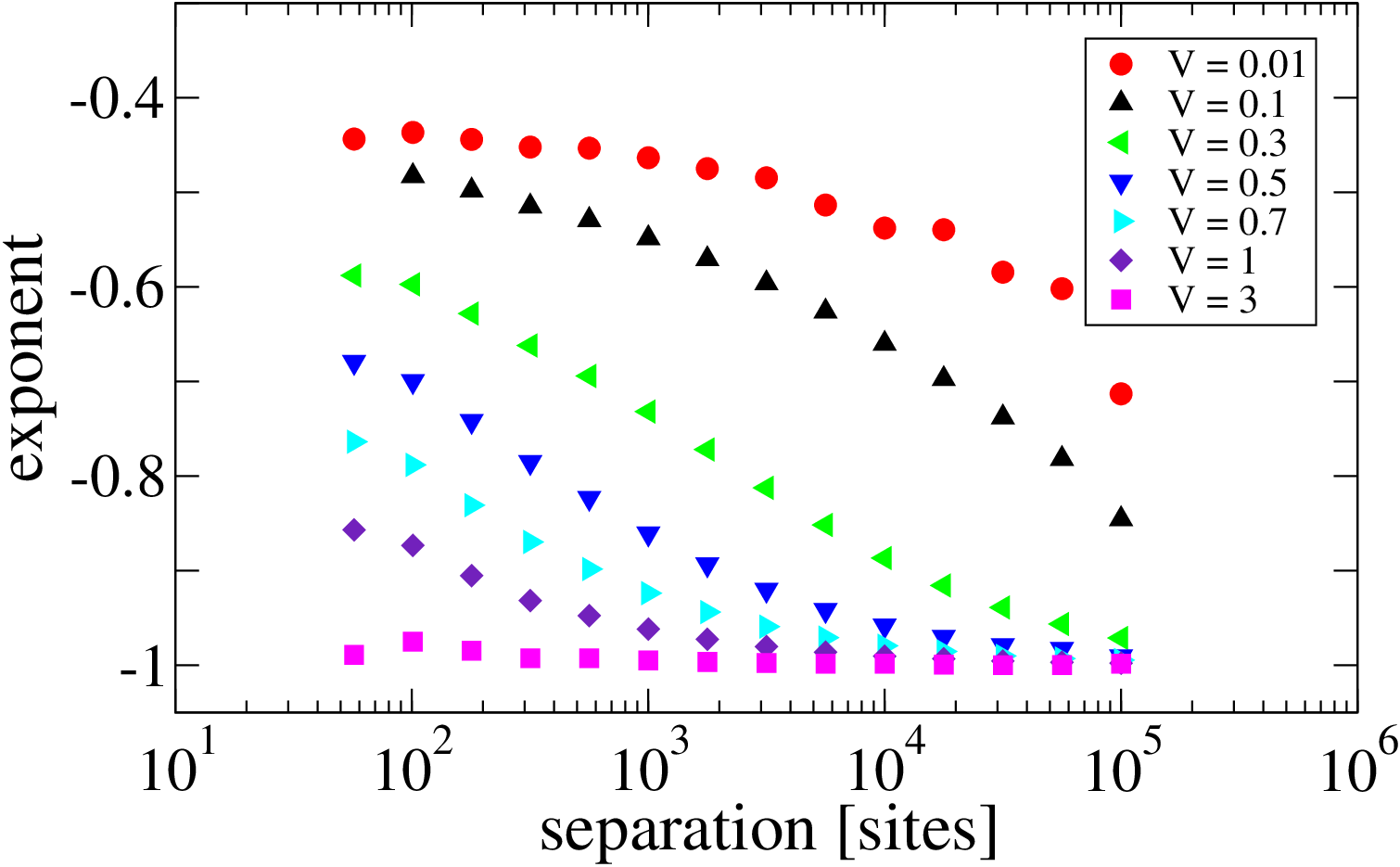}
\end{center}
\vspace{-0.6cm}
\caption[]{(Color online) Effective exponent of the amplitude of the
  indirect interaction for a 
collection of impurity strength $V_1=V_2=V$ and interaction $U=1$. The
``noise'' visible in the curve for $V=0.01$ originates from 
the fact that $V_{12}$ is very small for small $V$.\label{expWW1divVer}}
\begin{center}
\includegraphics[width=0.4\textwidth,clip]{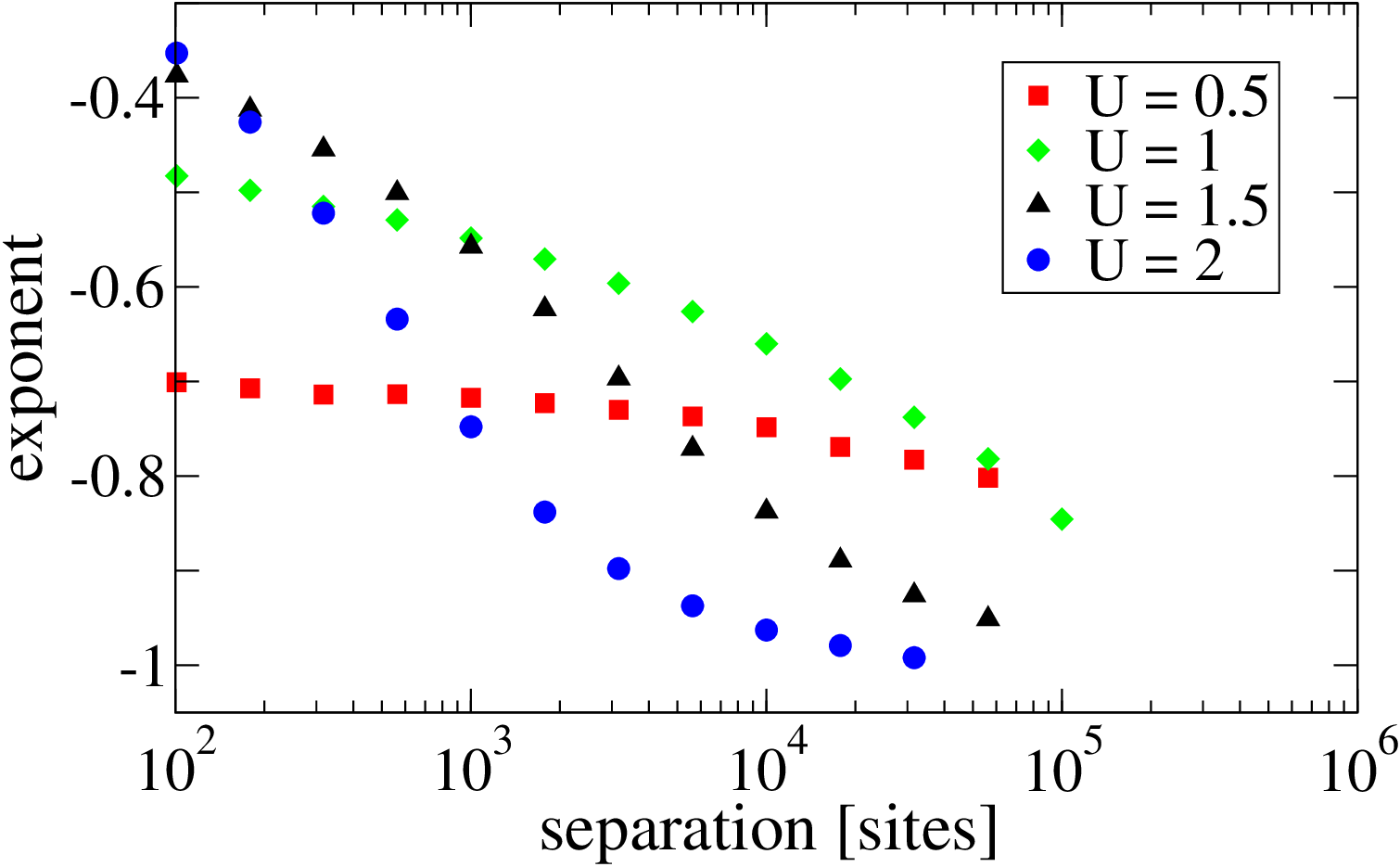}
\end{center}
\vspace{-0.6cm}
\caption[]{(Color online) The same as in Fig.\ \ref{expWW1divVer} but
  for fixed $V_1=V_2=0.1$ and different $U$.\label{expVp1divU}}
\end{figure}

For weak impurities $V_1$ and $V_2$ and separations $r$ 
sufficiently smaller then a crossover scale $r_c$ (see below), the 
indirect impurity interaction can again be understood using linear response 
theory. In 
this parameter regime the Friedel oscillations in the electron 
density induced by a single impurity behave as 
$(-1)^{r}/r^{2K-1}$.\cite{Egger95,Andergassen04} This 
leads to a potential of the same spatial dependence at the 
position of the second impurity.\cite{Recati05} 
This argument also holds for hopping impurities.
We thus expect 
that the oscillations of $V_{12}$ are symmetric and that the effective
exponent of the decay is $1-2K$, which corresponds to $-0.72$ 
for $U=0.5$ [see Eq.\
(\ref{exK})]. Our results are consistent with this argument (see Figs.\
\ref{WW1Vp1} and \ref{expWW1divVer}; in the latter figure $U=1$
leading to $1-2K=-0.5$). The small deviation of the
exponent at small $r$ can be
explained by our truncation of the exact hierarchy of fRG flow
equations. This leads to an approximate 
$K_{\rm fRG}$ which is slightly smaller than the exact $K$.\cite{Enss05}     

The linear response analysis
cannot be used at large $V_{1}$, $V_2$ and in analogy to 
the  $U=0$ considerations the appearance of the $1/r$ decay 
as well as the {\it oscillatory behavior} can be explained in terms of  
the finite size scaling of the ground state energy of a system 
with open boundary conditions. In lowest order perturbation theory in
$U$, $\Delta E$ is given by $-\pi (1+U/\pi)/(12 N)$ for even $N$. 
This is consistent with the (boundary) conformal field theory result 
Eq.~(\ref{bCFT}) as $v_c(U) = 2 (1+ U/\pi)$ to lowest order in $U$. 
For {\it odd} $N$ we did not succeed finding a closed form expression 
for the leading $U$, $1/N$ correction of the ground state 
energy. It is easy to show numerically that $\Delta E >0$ for odd $N$.
As in the noninteracting case this explains the repulsive (attractive) 
impurity interactions for site impurities with even (odd) separation and 
attractive (repulsive) $V_{12}$ for hopping impurities 
with odd (even) $r$. 
We note in passing that our numerics (lowest order perturabtion theory
as well as fRG) shows that for odd $N$, the $U$ dependence of 
$\Delta E$ cannot solely be expressed in terms of $v_c(U)$, as 
it is the case for even $N$. 

For $U>0$ linear response theory in addition breaks down for 
$r \gg r_c$ even if weak bare impurities are
considered. This is related to the fact that in a LL a single impurity is
a relevant perturbation in the renormalization group 
sense. Thus, for distances sufficiently larger then $r_c$, that is
in the low energy limit, even a single weak impurity acts as a
strong perturbation (flow towards the ``cut chain'' fixed 
point).\cite{LutherPeschel,Mattis,ApelRice,KaneFisher,Furusaki0} 
Consistent with this picture for $U>0$ and increasing $r$ we 
observe a tendency towards the exponent $-1$ in the decay of 
$V_{12}$ for weak to intermediate impurities 
(as in Figs.\ \ref{WW1Vp1} and \ref{expWW1divVer}).
Although the renormalization group flow  of the impurity
determines the {\it scaling exponents} and thus leads to the 
asymptotic $1/r$ decay of $V_{12}$, it does 
not affect the {\it sign} of the impurity interaction. The latter 
is fixed by the {\it bare} impurity strength 
(see Figs.\ \ref{WW1Vp1} and \ref{WW1V10}) and was discussed above.  
This observation shows that even for $r \gg r_c$ a system with weak 
to intermediate bare impurities is not completely equivalent to a 
system with strong bare impurities. 
 
In Fig.\ \ref{expWW1divVer} we show the effective exponent of 
the decay for interaction strength $U=1$ and various values of the
impurity strength $V_1=V_2=V$. The complete crossover from the weak 
to the strong impurity behavior of the decay exponent cannot be 
covered for a single $V$. This follows from the $V$ and $K$ 
(that is $U$) dependence of the single-impurity crossover 
scale\cite{KaneFisher}   
\begin{eqnarray}
\label{rcdef}
r_c \propto V^{1/(K-1)} \; .
\end{eqnarray}
For $V \ll 1$, $r_c$ grows exceedingly, even beyond the large 
system sizes we can treat using the truncated fRG.
The $K$ dependence of $r_c$ can be read off from Fig.\
\ref{expVp1divU},
where the effective exponent for $V_1=V_2=0.1$ and various values of
the interaction
strength is shown. The interaction ranges from $U=0.5$ to 
$U=2$, corresponding to LL parameters from $K \approx 0.86$ to 
$K\approx0.5$ [see Eq.\ (\ref{exK})]. For small to intermediate $U
\leq 1$ the effective exponent starts close to the linear response result 
$1-2K$ and tends towards the asymptotic value of $-1$ on a length 
scale decreasing with interaction strength. The asymptotic exponent
$-1$ is reached for separations beyond our system size. For larger
interactions $1 < U \leq 2$ already at the smallest separations
considered here a clear deviation from the linear response exponent is
observed, while the asymptotic exponent $-1$ is reached for $r \sim
10^5$.  

Similar to the noninteracting case in a narrow regime of {\it site} impurity 
parameters $V_1$ and $V_2$ the impurity interaction 
changes from being attractive (repulsive) for even
(odd) $r$ to the opposite. In the crossover region $V_{12}$ 
becomes purely attractive and one can again fine tune
$V_1$ and $V_2$ such that $V_{12}$ does not oscillate anymore.

\subsection{Attractive electron-electron interaction}

\begin{figure}[tb]
\begin{center}
\includegraphics[width=0.4\textwidth,clip]{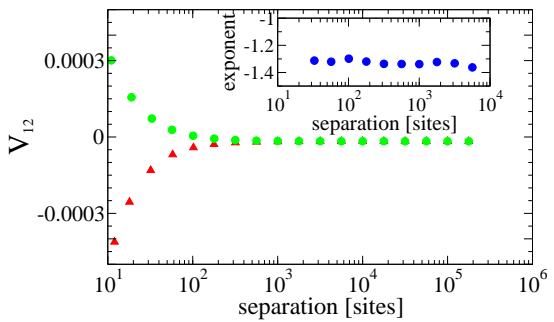}
\end{center}
\vspace{-0.6cm}
\caption[]{(Color online) Oscillations of the interaction energy as 
function of the separation for site impurities $V_1=V_2=1$ 
and {\it attractive} interaction $U=-0.5$. Triangles: 
even separations; circles: odd separations. 
Inset: exponent of the decay. The ``noise'' results from the small
values of $V_{12}$. \label{tt5WWm1V1}}
\end{figure}

Finally, we mention attractive interactions
$U<0$ (that is $K>1$) not considered so far. In this case 
a single impurity  is irrelevant in the renormalization 
group sense\cite{LutherPeschel,Mattis,ApelRice,KaneFisher,Furusaki0}   
which suggests that regardless of the strength of the bare impurities 
for $r \gg r_c$ linear response theory can be used.
We then anticipate that the impurity interaction oscillates 
around zero and for 
$r \gg r_c$ decays with the linear response exponent $1-2K$. 
Our numerical 
results are consistent with this expectation as exemplified in Fig.\
\ref{tt5WWm1V1} for site impurities with $V_1=V_2=1$ and $U=-0.5$,
leading to $K(U=-1) = 1.19$ and $1-2K=-1.38$. The small deviation in the
exponent again results from the approximate nature of our truncated
fRG equations.    
A similar behavior in the decay exponent of the Friedel 
oscillations induced by a single impurity was discussed in 
Ref.\ \onlinecite{Andergassen04} (see Fig.\ 15).
For $U<0$ accurate results for the exponent of the decay of $V_{12}$ 
are difficult to obtain as the impurity interaction becomes 
exceedingly small.

\section{Summary}
\label{summary}

To summarize, we examined 
the indirect interaction between two impurities placed in a 
LL. We started from a microscopic lattice
model, using the fRG in the interacting case. We 
observed, that for weak as well as strong impurities 
the interaction oscillates between being attractive and 
repulsive decaying algebraically with the separation of 
the impurities. In the 
case of noninteracting electrons the exponent of this 
decay is $-1$ independent of the impurity strength. For 
interacting electrons and weak impurity strength the powerlaw 
decay of $V_{12}$ starts at a larger $U$ dependent 
value $1-2K$ 
and tends towards $-1$ for large separations on a scale $r_c$  
depending on interaction and impurity strengths. 
For weak impurities this behavior can be understood
from the oscillatory decay of Friedel oscillations of the electron
density induced by an impurity using linear response theory. In 
the opposite limit the oscillatory $1/r$ decay can be traced back 
to the finite size scaling of the ground state energy of a chain 
with open boundary conditions. Our approach 
supplements analytic calculations obtained for large 
interactions\cite{Caux03} and those performed 
in the limits of large and small impurities.\cite{Recati05,Fuchs06} 
It reveals the full crossover from weak to
strong impurities. For attractive interactions the impurity
interaction oscillates and asymptotically (for large separations)
decays with the linear response exponent $1-2K$.  

\section*{Acknowledgments}
We thank J.N.\ Fuchs, A.\ Recati, 
and W.\ Zwerger for useful discussions. 
This work was supported by the Deutsche Forschungsgemeinschaft (SFB
602).


\begin{thebibliography}{99}
\bibitem{foot1} In the present model these are the impurity terms 
Eq.\ (\ref{hamimp}) for site impurities or Eq.\ (\ref{hamimpp}) for hopping
impurities and in addition the single-particle terms of 
$H_{\rm int}$ Eq.\ (\ref{hamint}).
\bibitem{fuchsreasoning} Within the reasoning of Ref.\
  \onlinecite{Fuchs06} this can be explained as follows: At
  half-filling and with hopping impurities our model is particle-hole
  symmetric. Thus forward scattering is forbidden by symmetry and the
  forward scattering phase shift $\delta$ vanishes for all $V_1$,
  $V_2$.
\end{thebibliography}
\end{document}